\documentclass[aps,prl,twocolumn,showpacs,superscriptaddress,floatfix]{revtex4-2}
\usepackage{dcolumn}
\usepackage{amsmath}
\usepackage{graphicx,epsfig,psfrag}
\usepackage{amssymb}
\usepackage{natbib}
\usepackage{url}
\usepackage[breaklinks=true,colorlinks=true,cit ecolor=blue,pdfusetitle]{hyperref}
\usepackage{mathtools}
\usepackage{subfigure}
\usepackage{bookmark}
\usepackage{epic}
\usepackage{txfonts}
\usepackage{color}
 \usepackage{soul}

\begin{document}

\title{Work distribution for strongly coupled many-body open quantum systems}

\author{H. T. M. Nghiem}
\affiliation{Phenikaa Institute for Advanced Study, Phenikaa University, 12116 Hanoi, Vietnam}
\author{T. A. Costi}
\affiliation
{Peter Gr\"{u}nberg Institut,
Research Centre J\"ulich, 52425 J\"ulich, Germany}
\author{S. Campbell}
\affiliation{School of Physics, University College Dublin, Belfield, Dublin 4, Ireland}
\affiliation{Centre for Quantum Engineering, Science, and Technology, University College Dublin, Ireland}
\author{A. K. Mitchell}
\affiliation{School of Physics, University College Dublin, Belfield, Dublin 4, Ireland}
\affiliation{Centre for Quantum Engineering, Science, and Technology, University College Dublin, Ireland}


\begin{abstract}
The non-equilibrium quantum thermodynamics of many-body open systems is notoriously rich, especially in the strong-coupling regime where strong system-bath correlations develop and interactions produce non-perturbative effects. Such systems can be described by `quantum impurity models' where the system and bath are treated on an equal footing as a single composite. Paradigmatic examples are the spin-boson and Anderson impurity models, in which system degrees of freedom interact with either bosonic or fermionic gapless baths. 
Here we study the quantum work distribution function (WDF) of such systems following a quench.  
Capturing the full continuum of many-body excitations in the WDF requires a non-perturbative solution of the underlying non-equilibrium quantum impurity problem. 
For this purpose, we extend the time-dependent numerical renormalization group (TDNRG) approach to the calculation of the WDF, which applies directly in the thermodynamic limit, can be used for arbitrary quench amplitudes at zero or finite temperature, and provides exponentially fine low-energy resolution. Our numerically-exact solution reveals power-law threshold behavior due to the Anderson orthogonality catastrophe when the work approaches its minimum value, with universal scaling collapse of the distribution below an emergent low-energy scale induced by system-bath correlations.
\end{abstract}


\maketitle
{\em Introduction.--}
Defining thermodynamic variables in a consistent manner for quantum systems has proven to be a subtle task, because quantum measurements are generally invasive and their outcomes stochastic~\cite{GemmerBook, Binder2019book, Landi2021,strasberg2021first,Gherardini2024}. 
An operational definition of the work done on a system due to an external drive can be formulated in terms of projective energy measurements before and after the driving protocol has been completed, which makes the work itself a stochastic variable~\cite{Talkner2007}. The probability \textit{distribution} of this quantum work then provides a detailed statistical characterization of the non-equilibrium process, and satisfies fundamental fluctuation theorems such as Jarzynski’s equality and the Crooks relation~\cite{Jarzynski1997,Crooks1999,batalhao2014experimental}. 
This has led to renewed interest in understanding the thermodynamics in situations where quantum effects cannot be neglected~\cite{campbell2026roadmap,miller2019work}. 

Several significant fundamental and practical advances have followed -- for example the exploration of deep connections between information and thermodynamics~\cite{sagawa2008second,toyabe2010experimental,masuyama2018information} and a sharpening of Landauer's bound~\cite{Reeb2014, aimet2025experimentally,de2025friendly}; the development of quantum thermal machines~\cite{peterson2019experimental,Koch2023, Aamir2025}; and proposals for quantum batteries~\cite{campaioli2024colloquium,quach2022superabsorption}. For many-body systems, work statistics under driving, particularly near quantum criticality, have been the focus of sustained interest~\cite{Silva2008, Dorner2012, Chenu2019, Fei2020,Ma2025}. 
Beyond the average work and its first moments, the full work distribution function (WDF) has emerged as a probe of quantum fluctuations, irreversibility, coherence, and many-body dynamics~\cite{Kiely2023, Santini2023, Zawadzki2020, Zawadzki2023, Solfanelli2025}. 

Despite this considerable progress, obtaining exact results remains extremely challenging for true many-body open quantum systems. 
Naturally, theoretical and computational studies have focused on aspects of the full problem in simpler limits. For example, key insights have been gained from processes close to equilibrium  -- either in the slow-driving regime~\cite{miller2019work,scandi2020quantum} or in the linear-response limit of weak driving~\cite{Ma2025,macieszczak2018unified}. Substantial simplifications also arise for non-interacting/independent particles or few-body problems~\cite{esposito2015quantum,bruch2016quantum,noman2026orthogonality}; and tractable systems with finite baths~\cite{pekola2016finite,riera2021quantum,Zawadzki2023,lacerda2023quantum}.

Although important results have been obtained using Markovian-type approximations~\cite{HekkingPekola2013,silaev2014lindblad,strasberg2016nonequilibrium}, systems that are strongly coupled to their environment can give more complex physics. Even in equilibrium, interactions produce many-body correlations between system and bath and emergent low-energy scales. Such cases can be described by `quantum impurity models' in which the system and bath are treated together as a single interacting composite, allowing arbitrary system-bath correlations~\cite{Hewson1997}. The `impurity' system 
induces a non-trivial renormalization group (RG) flow and universal scaling behavior at low energies and temperatures~\cite{Leggett1987,KWW1980a,affleck1991kondo,Bulla2008}.
The simplest examples consist of a single qubit impurity interacting with a gapless non-interacting bath of either bosons or fermions, characterized by a continuous spectral density in the thermodynamic limit. Respectively, these are the celebrated spin-boson~\cite{Leggett1987} and Kondo~\cite{Hewson1997} models. Interestingly both models lack a weak-coupling limit, 
meaning that the physics is non-perturbative: even for very weak bare system-bath coupling, the \textit{effective} coupling at low energies becomes strong under RG (a similar mechanism underpins the asymptotic freedom phenomenon in particle physics). 

To understand the properties of the full WDF of such systems following a drive, one must therefore solve the underlying non-equilibrium impurity model non-perturbatively to capture the strong-correlation physics and resolve the universal many-body continuum of work processes at low energies. 

In this Letter we develop the time-dependent numerical renormalization group (TDNRG)  method~\cite{Anders2005,Anders2006,Nghiem2014a,Nghiem2014b} to compute the WDF of generalized quantum impurity models following a quench. TDNRG is a non-perturbative technique, based on a tensor-network reformulation of Wilson's numerical RG procedure~\cite{Bulla2008}, which applies in the thermodynamic limit and can be used for arbitrary quench amplitudes. It provides
exponentially fine resolution of the WDF at low energies and can access the universal long-time, low-temperature regime.

\begin{figure}[t]
    \centering
    \includegraphics[width=0.48\textwidth]{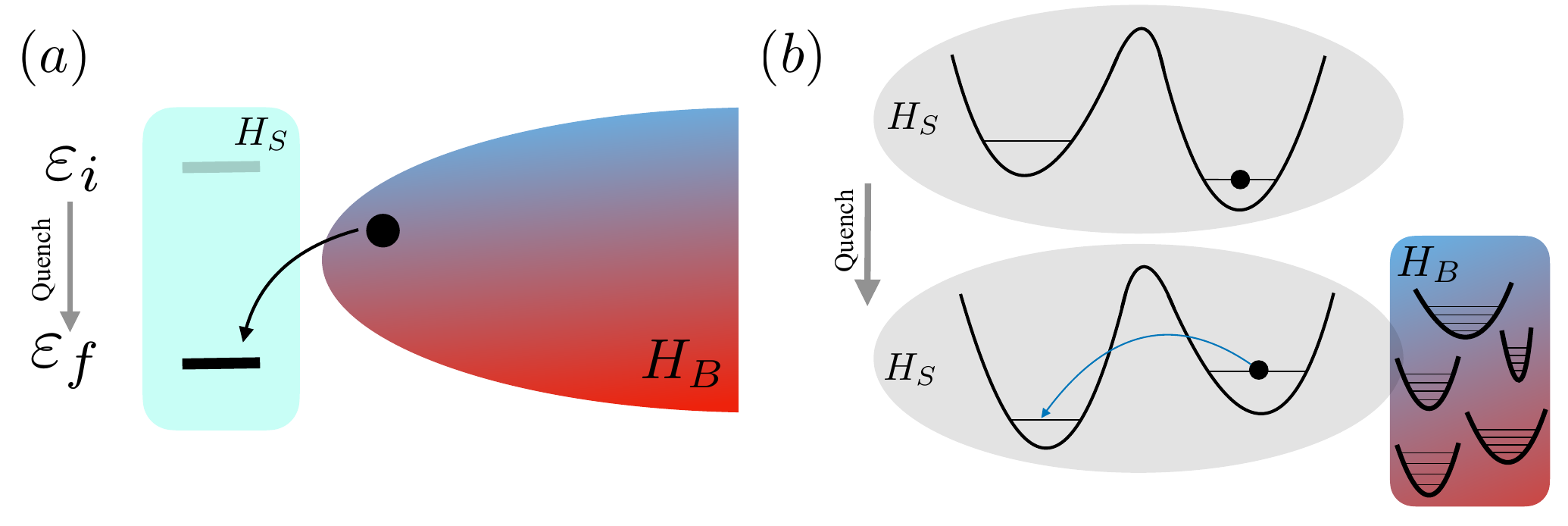}
    \caption{Schematic illustration of the systems and quench protocols. (a) Anderson impurity model describing a single-level interacting quantum dot coupled to an electronic lead.  (b) Ohmic spin-boson model describing dissipative barrier tunneling. Work is done by suddenly changing the dot potential or biasing the tunnel barrier states.}
    \label{fig:schematic}
\end{figure}

In the following we determine the full WDF due to a sudden quench  in two paradigmatic quantum impurity models, and interpret these in terms of Anderson's orthogonality catastrophe (OC). For both models, the WDF at $T=0$ follows a robust power-law threshold behavior as the work approaches its minimum, with exponents determined by the OC and universal scaling below an emergent (interaction-renormalized) work value. We also validate our approach by verifying the Crooks fluctuation theorem at finite $T$. Further technical details and analysis appear in a companion paper~\cite{Nghiem2026prb}.


{\em Quench protocol and orthogonality catastrophe.--}
A quench involves changing system parameters suddenly. Although finite-time driving allows a wider range of physics to be explored~\cite{Ma2025,lotem2026multibath}, the sudden-quench limit is in fact practically relevant in many physical settings and still gives nontrivial thermodynamic behavior~\cite{Tureci2011, Kasztelan2011, Heyl2012a, Torre2013, Marino2014, Campbell2016, Lobejko2017, Bernien2017, Mishra2018, Davoudi2025}. A dramatic example of this is the OC scenario~\cite{Anderson1967, Nozieres1969}:  after a quench, the bath states associated with the pre- and post-quench system configurations become orthogonal in the thermodynamic limit, even though the perturbation was local. A change in the system therefore produces a collective, infrared-singular rearrangement of the environment. Since infinitely many environmental degrees of freedom ``record'' the state of the system and cannot respond to the change instantaneously, local driving is met with a kind of quantum drag, beyond simple Markovian dynamics. 

While several works have considered the thermodynamic implications of this in various manifestations of the OC~\cite{Goold2011, Heyl2012b, Sindona2013, Fogarty2020}, a full analysis for many-body open systems in the strong coupling regime has not previously been possible.


{\em Quantum work distribution.--}
The stochastic quantum work $W$ is here defined through a two-time projective energy measurement before and after the quench, and is distributed according to the WDF, $P(W)$. With $H_i$ and $H_f$ the initial and final (pre- and post-quench) Hamiltonians satisfying  $H_{i/f}|n_{i/f}\rangle = E_{n_{i/f}}|n_{i/f}\rangle$, the WDF 
is given by~\cite{Talkner2007},
\begin{eqnarray}
P(W)= \sum_{m,n}
p_{n_i}
\left|\langle m_f|n_i\rangle\right|^2
\delta\!\left(
W-E_{m_f}+E_{n_i}\right)\;,
\end{eqnarray}
where $p_{n_i}=e^{-\beta E_{n_i}}/Z_i$ and $Z_i=\sum_m e^{-\beta E_{m_i}}$ for an initial thermal state. The WDF can alternatively be obtained from the Fourier transform of the characteristic function $G(u) = \langle e^{iuH_f}e^{-iuH_i}\rangle$. 


{\em Models and physical systems.--} 
The quantum impurity models we consider are of the form $H=H_{S}+H_{int}+H_{B}$ consisting of a small quantum mechanical system ($H_S$) coupled to an infinite environment ($H_B$) via a system-bath interaction ($H_{int}$). Here we focus on two of the simplest such models (which nevertheless exhibit non-trivial many-body physics) but note that our methodology can be applied to more complex systems. 

The Anderson impurity model (AIM)~\cite{Anderson1961,Hewson1997} involves a single spinful interacting fermionic site coupled to a metallic bath of conduction electrons, see Fig.~\ref{fig:schematic}a. Here $H_{S}=\sum_{\sigma}\varepsilon_d n_{d\sigma} + U n_{d\uparrow}n_{d\downarrow}$ describes the impurity with level potential $\varepsilon_d$ and Coulomb repulsion $U$. The fermionic bath is given by $H_{B}=\sum_{k\sigma}\varepsilon_{k}c_{k\sigma}^{\dagger}c_{k\sigma}$ and $H_{int}=V\sum_{k\sigma}( c_{k\sigma}^{\dagger}d_{\sigma}+d_{\sigma}^{\dagger}c_{k\sigma})$ couples the system and bath. The hybridization strength is $\Gamma=\pi\rho V^2$ where $\rho=1/2D$ is a constant conduction density of states with half bandwidth $D=1$. 
Near half-filling, the AIM maps onto the Kondo model~\cite{Hewson1997}; at equilibrium the impurity spin is dynamically screened by the Kondo effect below a low-energy emergent scale $T_K$ through formation of a many-body singlet state inside a large entanglement cloud of bath electrons~\cite{lee2015macroscopic}.  
The AIM is the effective model describing semiconductor quantum dot circuits~\cite{goldhaber1998kondo,cronenwett1998tunable}. In such devices the dot level $\varepsilon_d$ can be manipulated \textit{in-situ} with a gate voltage.

The second setting we consider is the famous spin-boson model (SBM), describing a qubit coupled to a bosonic bath of harmonic oscillators. Physically, it captures dissipative tunnelling between two localized states, for example a particle moving between the minima of a double-well potential, see Fig.~\ref{fig:schematic}b. Here $H_S=-\frac{1}{2}\Delta_0\sigma_x + \frac{1}{2}\epsilon\sigma_z$ describes the two-level system with bare tunneling amplitude $\Delta_0$, subject to a bias $\epsilon$. The bosonic bath is given by $H_B=\sum_{i}\omega_i(a_{i}^{\dagger}a_{i}+1/2)$, with $0\leq\omega_i\leq \omega_c$ and the interaction term is $H_{int}=\frac{1}{2}\sigma_{z}\sum_{i}\lambda_{i}(a_{i}+a_{i}^{\dagger})$. We consider an Ohmic bath characterized by the linear spectral density $J(\omega)\equiv\pi\sum_{i}\lambda_{i}^2\delta(\omega-\omega_i)=2\pi\alpha\omega$. The dimensionless quantity $\alpha$ encodes the dissipation strength. At equilibrium, coupling to the bath strongly renormalizes the tunneling amplitude, generating an emergent low-energy scale $\Delta_r$. For the Ohmic model, tunneling is progressively suppressed with increasing dissipation
strength $\alpha$, culminating in a localization transition at $\alpha=1$. Below $\Delta_r$, the qubit is strongly dressed by a collective cloud of low-energy bath excitations. As discussed in the \textit{End Matter}, we employ a mapping~\cite{Guinea1985,Nghiem2016} to the equivalent spinless interacting resonant level model (IRLM), which can be treated more efficiently within TDNRG.

In the following we consider level quenches in the AIM, $\varepsilon_d=\varepsilon_d^i\to \varepsilon_d^f$ with the interaction $U$ fixed. In the SBM we examine quenches from fully biased $\epsilon=\epsilon^i=-\infty $ to unbiased $\epsilon^f=0$ at fixed dissipation strength $\alpha$. Other scenarios are discussed in the companion paper~\cite{Nghiem2026prb}.


{\em TDNRG method.--}
We evaluate the characteristic function $G(u)$ and hence obtain the work distribution $P(W)$ using the complete-basis formulation of TDNRG~\cite{Anders2005,Anders2006,Nghiem2014a,Nghiem2014b}.
Within NRG, the continuum bath is logarithmically discretized and
mapped onto a Wilson chain whose couplings decrease exponentially
along the chain.  For fermionic Wilson chains, 
the characteristic energy scale at iteration $m$ behaves as
$\omega_m\sim D\Lambda^{-m/2}$, where $\Lambda>1$ is the NRG
discretization parameter.  The Hamiltonian is diagonalized
iteratively as successive Wilson sites are added, retaining a set of
low-energy states ($K$) and discarding the remaining high-energy
states ($D$).  We perform separate NRG calculations for the initial
and final Hamiltonians, $H_i$ and $H_f$, obtaining shell eigenstates
$|rm\rangle_{i/f}$ with energies $E_r^{m,i/f}$. Here $r$ labels different states at the same $m$.

A complete many-body basis for the Wilson chain can be constructed
from the discarded states of all NRG iterations
\cite{Anders2005}.  Specifically, a state $|lm\rangle_X$
discarded at iteration $m$ is supplemented by an environment state
$|e\rangle$ describing the remaining Wilson sites $m+1,\ldots,N$,
such that $|lme\rangle_X=|lm\rangle_X\otimes|e\rangle$, with
$X=i,f$.  These states obey the completeness relation
\begin{eqnarray}
\mathbf{1}
=
\sum_{m=m_0}^{N}\sum_{l,e}
|lme\rangle_X\,{}_X\langle lme| ,
\qquad X=i,f ,
\label{eq:cbs}
\end{eqnarray}
where $m_0$ is the first iteration at which truncation is performed,
and all states at the final iteration $N$ are regarded as discarded.
Within the standard NRG approximation, energy-scale separation allows
one to associate each such state with the shell energy,
$H_X|lme\rangle_X\simeq E_l^{m,X}|lme\rangle_X$; the much lower-energy
environment degrees of freedom therefore do not enter this energy.

For an initial thermal state
$\rho_i=e^{-\beta H_i}/Z_i$, we use the full-density-matrix NRG
construction, which incorporates thermal contributions from all
Wilson shells \cite{Weichselbaum2007,Nghiem2014a}.
For the local quenches considered here the environment basis is common
to the initial and final Hamiltonians, and their many-body overlaps
therefore factorize as
\begin{eqnarray}
{}_f\langle rme|qme'\rangle_i
=
S^m_{rq}\,\delta_{ee'} ,
\qquad
\widetilde{R}^{m}_{qs}
=
\sum_e {}_i\langle qme|\rho_i|sme\rangle_i ,
\label{eq:overlap_rdm}
\end{eqnarray}
where $S^m_{rq}={}_f\langle rm|qm\rangle_i$ is the overlap between
final- and initial-state NRG eigenbases, while
$\widetilde{R}^{m}$ is the initial full density matrix reduced over
the environment sites beyond iteration $m$. 
Inserting the complete NRG bases into
$G(u)=\mathrm{Tr}\!\left[e^{iuH_f}e^{-iuH_i}\rho_i\right]$
and using the NRG approximation then gives
\begin{eqnarray}
G(u)
=
\sum_{m=m_0}^{N}
\sum_{r,s,q}^{\prime}
S^m_{rq}\,
\widetilde{R}^{m}_{qs}\,
\left(S^m_{rs}\right)^{*}
e^{iu(E_r^{m,f}-E_q^{m,i})}.
\label{eq:G_TDNRG}
\end{eqnarray}
The primed sum is over the kept and discarded sectors of the three
states $r,s,q$ at iteration $m$, excluding only the case in which all
three are kept.  The latter contribution is represented at subsequent
NRG iterations and must therefore be omitted to avoid double counting.
This converts the three independent complete-basis sums into a single
sum over Wilson shells.

Finally, Fourier transforming Eq.~\eqref{eq:G_TDNRG} yields the WDF,
\begin{eqnarray}
P(W) = \sum_{m=m_0}^{N}
\sum_{r,s,q}^{\prime}
S^m_{rq}\,
\widetilde{R}^{m}_{qs}\,
\left(S^m_{rs}\right)^{*}
\delta\!\left[
W-\left(E_r^{m,f}-E_q^{m,i}\right)
\right].
\label{eq:P_TDNRG}
\end{eqnarray}
Thus each Wilson shell contributes transitions at its characteristic
energy scale, with the initial thermal weights contained in
$\widetilde{R}^{m}$ and the effect of the quench encoded in the
initial--final overlap matrices.  The exponentially decreasing Wilson energy scale provides correspondingly fine resolution of $P(W)$ close to threshold.  Completeness of the NRG basis gives $G(0)=1$ and hence $\int dW\,P(W)=1$.  The resulting discrete $\delta$-peak spectrum is broadened only when constructing the smooth WDF shown below. A full derivation and discussion is provided in the companion paper, Ref.~\cite{Nghiem2026prb}.

Our TDNRG method thus brings the non-perturbative calculation of the WDF within reach for the class of many-body, strongly coupled open systems described by quantum impurity models. As shown below, we find that the WDF exhibits universal scaling of low-energy many-body excitations that form an edge singularity, encoding the impurity RG flow.


\begin{figure}[t]
    \centering
    \includegraphics[width=0.48\textwidth]    {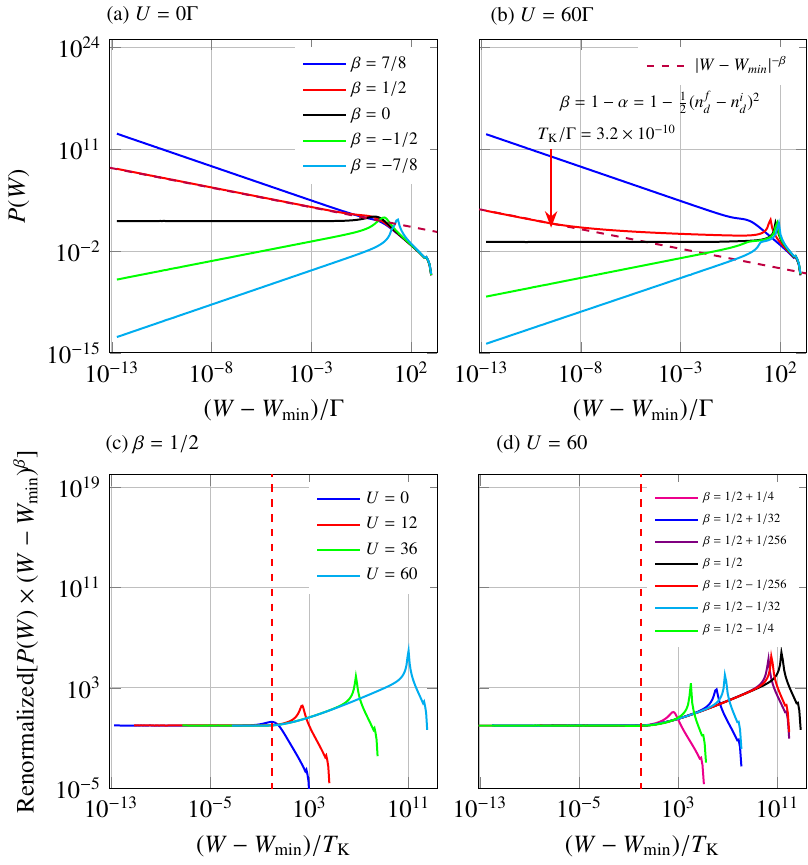}
    \caption{WDFs for level quenches with $U=0$ are shown in (a) and with finite $U$ in (b). The red arrow in (b) indicates the final-state Kondo temperature of the $\beta=1/2$ system (whose final state is in the Kondo regime, $n_f=1$). (c,d) WDF $P(W)\times (W-W_{\textrm{min}})^{\beta}$ normalized to its $W=W_{\rm min}$ value. Scaling (data collapse) is observed below $W-W_{\rm min}=T_{\rm K}$ for fixed $\beta=1/2$ and varying $U$ in (c), and, for fixed $U=60\Gamma$ and varying $\beta$ in (d).}
    \label{fig:AOC}
\end{figure}

{\em WDF of the quenched AIM.--}
Noting the connection between WDFs and response functions in the x-ray problem~\cite{Heyl2012a,Heyl2012b}, we expect that at zero temperature, $P(W)$ will, as in the case of spectral functions \cite{Costi1994,Costi1996b,Helmes2005}, be dominated by orthogonality effects between the initial and final groundstates, resulting in the asymptotic form $P(W\to W_{\rm min})\sim |W-W_{\rm min}|^{-(1-\alpha_{\rm OC})}$ with $\alpha_{\rm OC}=\sum_{\sigma}(\Delta \delta_{\sigma}/\pi)^2$ denoting the Anderson orthogonality exponent and $\Delta\delta_\sigma=\delta_\sigma^i-\delta_\sigma^f$ being the difference between initial and final state phase shifts. The minimum work, $W_{\rm min}$, given by the difference between initial and final groundstate energies, acts as the threshold energy in the x-ray problem, with $P(W)=0$ for $W<W_{\rm min}$ at $T=0$. From the Friedel sum rule for the AIM \cite{Hewson1997}, we have that $\delta_{\sigma}^{i,f}=\pi n_{d\sigma}^{i,f}$, where $n_{d\sigma}^{i,f}$ are the spin resolved initial/final state occupations of the local level. Hence, we expect a power law behavior $P(W)\sim |W-W_{\rm min}|^{-\beta}$ with exponent $\beta=1-(n_f-n_i)^2/2$ where $n_{i,f}=\sum_{\sigma} n_{d\sigma}^{i,f}$.

We consider level quenches such that the occupation in the initial state is zero (setting $\varepsilon_d^i=D$), while $\varepsilon_d^f$ is chosen such that the occupation in the final state satisfies a given chosen value of $\beta$.  Figures~\ref{fig:AOC}(a)-\ref{fig:AOC}(b) show the convergent and divergent WDFs for the noninteracting [Fig.~\ref{fig:AOC}(a)] and interacting [Fig.~\ref{fig:AOC}(b)] cases, respectively, for 
various values of $\beta$. 

\begin{figure}[t]
    \centering
    \includegraphics[width=0.48\textwidth]{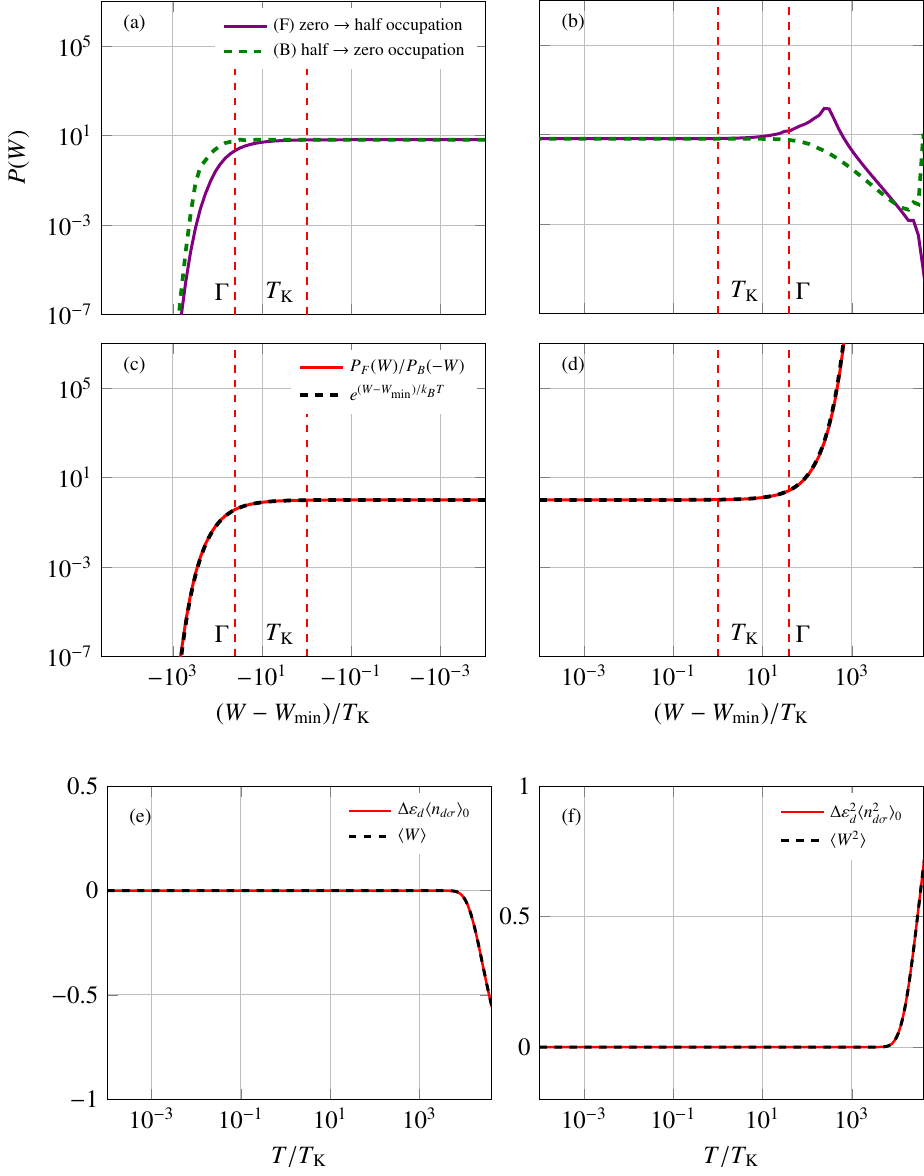}
     \caption{TDNRG results for the work at finite temperatures for a level quench in the AIM with $U=12\Gamma$ and $\beta=1/2$. WDFs for the forward quench (purple line) and the backward quench (green line) at $T=\Gamma$ are shown in (a) for the negative range of $W-W_{\textrm{min}}$; and in (b) the positive range of $W-W_{\textrm{min}}$ . The corresponding ratio of the forward-quench WDF and the backward-quench WDF (red line) and the Crooks relation (black line) is shown in (c) and (d). The first and second moments of the WDF are shown in (e) and (f).}
    \label{fig:crooks}
\end{figure}

The scale for the onset of the power law in $P(W)$ is the low energy scale of the final state. This will be the Kondo scale, $T_{\rm K}=\sqrt{\Gamma U^f/2} \exp(\pi \varepsilon_d^f (\varepsilon_d^f+U^f)/2\Gamma U^f)$, when the final state is in the Kondo regime \footnote{$T_{\rm K}=\Gamma$ in the mixed valence regimes $\beta=7/8,0$ with $n_f=1/2,\sqrt{2}\approx 1.4$ and $T_{\rm K}=\varepsilon_f$ in the full orbital regimes $\beta=-1/2,-7/8$ ($n_f=\sqrt{3}\approx 1.7,\sqrt{15/4}\approx 1.9$) }. We therefore expect universal scaling for $W - W_{\rm min} < T_{\rm K}$. This is clearly seen for the case $\beta=1/2$ and $U=60\Gamma$ in Fig.~\ref{fig:AOC}(b). For this quench, the final state is close to the symmetric Kondo regime ($n_f\approx 1$) and 
$T_{\rm K}$ is exponentially small at $U_f=60\Gamma$. 
Since the WDFs follow asymptotically a power law at $W - W_{\rm min} < T_{\rm K}$, the WDFs $P(W)\times (W-W_{\textrm{min}})^{\beta}$ approach a constant $c(\beta,U)$ that depends on $\beta$ and $U$ at lowest $W$. Normalizing by $c(\beta,U)$ results in scaling collapse of the WDF below $W - W_{\rm min} < T_{\rm K}$, as shown in Fig.~\ref{fig:AOC}(c) for the case of fixed $\beta=1/2$ with varying values of $U$, and in Fig.~\ref{fig:AOC}(d) for the case of fixed $U$ upon varying  $\beta$. 

\begin{figure}[t]
    \centering
    \includegraphics[width=0.48\textwidth]{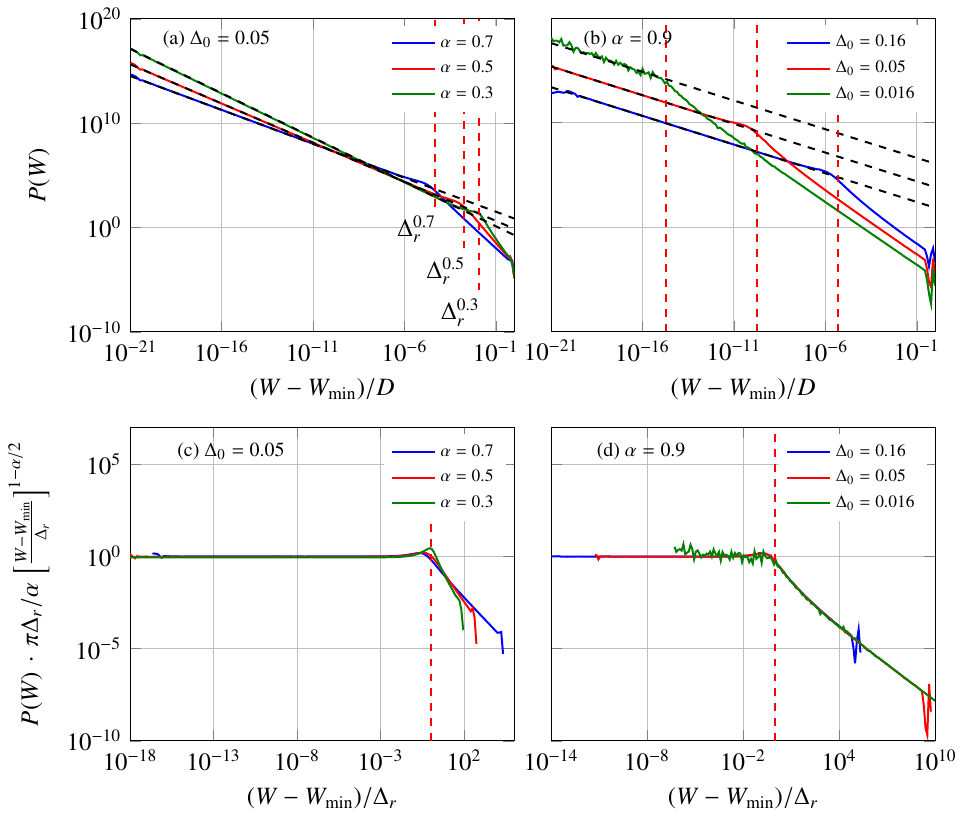}
    \caption{Work distribution for quenches in the SBM. 
    Top panels: the WDF $P(W)$ at $T=0$. (a) Bias quench $\epsilon_d=-\infty \to 0$ for several $\alpha=\alpha_i=\alpha_f<1$ and fixed $\Delta_0=0.05$. (b) Fixed $\alpha_i=\alpha_f=\alpha=0.9$ for several $\Delta_0$. Analytical asymptotes $P(W)\sim |W-W_{\rm min}|^{-(1-\alpha/2)}$ shown as black lines. Vertical red lines indicate the final state value of $\Delta_r$. Lower panels (c,d): rescaled curves exhibiting data collapse. See Table~\ref{tab:parameters} for the corresponding IRLM parameters.
    }
    \label{fig:work_IRLM_quench}
\end{figure}

At finite temperature, the validity of the numerical calculations can be checked by verifying the Crooks relation \cite{Crooks1999}, which connects the finite temperature WDFs for forward (F) and backward (B) quenches, via $\frac{P_F(W+W^{F}_{\rm min})}{P_B(-W+W^{B}_{\rm min})}=e^{\frac{W}{k_{\rm B}T}}$. We check the relation for the case of a level quench with $U=12\Gamma$ and $\beta=1/2$. Figs~\ref{fig:crooks}(a,b) show the WDFs of the forward quench (purple line) and the backward quench (green line) at finite temperature $T=\Gamma$ in the negative and positive range of $W-W_{\rm{min}}$ respectively. At finite $T$ the WDFs do not follow the power law at low $W$, as observed at zero temperature in Fig.~\ref{fig:AOC}(a,b), but flatten out for $|W-W_{\textrm{min}}|<T$. Figs.~\ref{fig:crooks}(c,d) show the ratio of the forward-quench WDF and the backward-quench WDF (red line), compared with the factor $e^{\frac{W}{k_{\rm B}T}}$ (black line), with the perfect agreement indicating that the Crooks relation is exactly satisfied. An additional check on the numerical calculations is obtained by checking the moments of the WDFs. The zeroth moment is satisfied exactly (probability conservation), while the first-order and second-order moments can be evaluated analytically~\cite{Ma2025}, being given by $\langle W\rangle=\Delta\varepsilon_d \langle n_d\rangle_0$ and $\langle W^2\rangle=\Delta\varepsilon_d^2 \langle n_d^2\rangle_0$ for a level quench, in which $\Delta\varepsilon_d$ is the quench size and $\langle ...\rangle_0$ is the expectation value in the initial state. The numerically evaluated moments from $P(W)$ agree well with the analytical ones over a wide range of temperatures as shown in Fig.~\ref{fig:crooks}(e,f). This establishes the high accuracy and resolution of our TDNRG calculation over exponentially separated work scales.


{\em WDF of the quenched SBM.--} Finally, we study the WDF for the Ohmic SBM subject to a bias quench. Specifically, here we take quenches from a fully polarized initial state to an unbiased final state ($\epsilon=\epsilon_i =-\infty\to \epsilon_f=0$). TDNRG results at $T=0$ are shown in Fig.~\ref{fig:work_IRLM_quench}(a) for several dissipation strengths $\alpha=\alpha_i=\alpha_f<1$ in the tunneling regime and for fixed $\Delta_0$. Fig.~\ref{fig:work_IRLM_quench}(b) shows results at fixed dissipation strength $0<\alpha=\alpha_i=\alpha_f<1$ in the tunneling regime for several $\Delta_0$.

The observed threshold behavior of $P(W)=(W-W_{\rm min})^{-\beta}$ for $W-W_{\rm min}\ll \Delta_r(\alpha)$ with $\beta=1-\alpha_{{\rm OC}}$ can be understood in terms of the OC effect, where $\alpha_{{\rm OC}}$
is the orthogonality exponent. Within the equivalent IRLM, this exponent is determined as the difference between the initial and final state phase shifts  $\alpha_{{\rm OC}}=(\Delta\delta/\pi)^2$ and is related to the dissipation strength via $\alpha_{\rm OC}=\alpha/2$, see \textit{End Matter}. The resulting $P(W)=(W-W_{\rm min})^{-\beta}$ with $\beta=1-\alpha_{\rm OC}=1-\alpha/2$ is consistent with the numerical calculations shown in  Fig.~\ref{fig:work_IRLM_quench}~(a,b). The asymptotic approach to the power law behavior below the emergent renormalized low energy scale $\Delta_r$ allows data collapse for $P(W)$ at $W-W_{\rm min}\ll \Delta_r(\alpha)$ for different $\alpha$ as seen in Fig.~\ref{fig:work_IRLM_quench}(c). Similarly, the universality of $P(W)$ with respect to the single low energy scale $\Delta_r(\alpha)$ at fixed $\alpha$ and different bare tunneling amplitudes $\Delta_0$ is demonstrated in Fig.~\ref {fig:work_IRLM_quench}(d). Thus, for the bias quench considered, the dissipation strength is seen to fully characterize the threshold behavior of the work distribution function in the SBM, while universal data scaling collapse emerges only below $\Delta_r$


{\em Conclusions and Outlook.--} We have generalized the TDNRG approach to obtain the full quantum work distribution of strongly coupled many-body open systems following a sudden quench, treating the system and its environment non-perturbatively as a single quantum impurity problem in the thermodynamic limit.  For both the fermionic AIM and bosonic SBM, we find that the low-work WDF at $T=0$ is governed by the Anderson orthogonality catastrophe: the continuum develops a universal power-law threshold whose exponent is fixed by the change in low-energy scattering phase shifts.  Interactions enter not only through this exponent, but also through the emergent scale below which the universal regime is reached, namely $T_K$ for the Kondo problem, and $\Delta_r$ for the dissipative two-level system.  The WDF therefore provides a direct thermodynamic probe of the impurity RG flow, resolving both its infrared fixed-point structure and the interaction-renormalized crossover scale.  At finite temperature we have further verified the formalism through the Crooks fluctuation relation and exact moment sum rules.

A particular strength of TDNRG in this setting is its logarithmic
separation of energy scales, which provides exponentially fine
resolution of the WDF close to threshold and allows exponentially
small many-body scales to be accessed directly.  This makes the
approach complementary to recent Monte Carlo and real-time
process-tensor methods~\cite{zhao2025monte,popovic2021quantum,shubrook2025numerically,lotem2026multibath}, for which resolving the asymptotic infrared
regime at low temperatures can be particularly demanding.  More generally, the present framework opens the door to work statistics for richer quantum impurity problems, including multichannel and multiorbital systems and quenches involving interaction or system-bath coupling parameters.  An important further direction is finite-time driving, which may be constructed within TDNRG from sequences of quenches~\cite{Nghiem2014b}, allowing the evolution of the universal threshold physics away from the sudden limit to be explored.  A detailed derivation of the formalism, additional benchmarks, and a broader range of quench protocols are presented in the companion paper~\cite{Nghiem2026prb}.


\begin{acknowledgments}
\textit{Acknowledgments.--}
H.T.M.N.~is funded by Vietnam National Foundation for Science and Technology Development (NAFOSTED) under grant number 103.02-2021.95. S.C.~acknowledges support from Taighde \'Eireann - Research Ireland under Grant No.~24/EPSRC/4121. 
A.K.M.~acknowledges support from Taighde \'Eireann - Research Ireland under Grant No.~24/FFP-P/12816. S.C.~and A.K.M. acknowledge financial support of Taighde \'Eireann – Research Ireland, under Grant No.~23/RC/12197 at Rinn Quantum. We acknowledge computing time on the supercomputer JURECA \cite{JURECA} at Forschungszentrum J\"ulich under grant no.~JIFF23 and the PHENIKAA University's HPC Systems. 


\appendix
\section*{End Matter}
\label{sec:endmatter}

\setcounter{figure}{0}
\renewcommand{\thefigure}{E\arabic{figure}}

\setcounter{equation}{0}
\renewcommand{\theequation}{E.\arabic{equation}}

\setcounter{table}{0}
\renewcommand{\thetable}{E.\arabic{table}}

\setcounter{section}{0}
\renewcommand{\thesection}{E.\Roman{section}}

\renewcommand{\thesubsection}{E.\Roman{section}.\Alph{subsection}}
\makeatletter
\renewcommand*{\p@subsubsection}{} 
\makeatother

\textit{Interacting resonant level model and equivalence to the Ohmic spin-boson model.--}
\label{sec:appendix-IRLM}
In this Letter, the results for the Ohmic spin-boson model are obtained by exploiting the equivalence of this model to the
spinless interacting resonant level model (IRLM) and applying the TDNRG approach to the latter. The IRLM, defined by $H=H_S+H_B+H_{int}$, consists of a system part
$H_{S}=\epsilon_d(d^{\dagger}d-dd^{\dagger})+V_0(c_{0}^\dagger d + d^\dagger c_{0})$, describing a two level system with tunneling amplitude $V_0$ and subject to a bias $2\epsilon_d$, interacting via a non-local Coulomb interaction $H_{int}=U_{dc}(n_d-1/2)(n_0-1/2)$ with a bath of spinless conduction electrons $H_B=\sum_{k}\epsilon_{k}c_{k}^\dagger c_{k}$. The correspondence to the SBM reads $\Delta_0=2V_0, \epsilon=2\epsilon_d$ and $\alpha = \frac{1}{2}(1+\frac{2\delta}{\pi})^2$ with phase shift $\delta=\arctan(-\pi\rho U_{dc}/2)$ and constant density of states $\rho=1/2D=1/\omega_c$. The dissipation strength $\alpha$ ranges from $1/2$ to $0$ as $U_{dc}$ is varied from $0$ to infinity (coherent tunneling regime), from $1/2$ to $1$ as $U_{dc}$ is varied from $0$ to $U_{dc}^{*}=-(2/\pi\rho)\tan(\pi(\sqrt{2}-1)/2)\approx -0.969$ (incoherent tunneling regime) and from $1$ to $2$ as $U_{dc}$ is reduced from $U_{dc}^*\approx -0.969$ to $-\infty$ (localized regime). The quenches 
studied in this Letter are all within the tunneling regime $0\leq \alpha < 1$ ($+\infty \geq U_{dc}>U_{dc}^{*}$). In this regime, the low energy scale of the SBM, the renormalized tunneling amplitude, $\Delta_r$, given approximately by $\Delta_r/\omega_c \sim (\Delta_0/\omega_c)^{1/(1-\alpha)}$, can be identified with the low energy scale, $T_0$, of the IRLM . The latter is given by $T_0=1/2\chi(0)\equiv \Delta_r$,  where $\chi(0)=-\frac{1}{2}\partial n_d/\partial \epsilon_d$ is the zero temperature local charge susceptibility of the IRLM. Table~\ref{tab:parameters} lists typical values of $\alpha$, $\Delta_r=T_0$ and the threshold exponent $\beta$ in the WDF as a function of $U_{dc}$.

\textit{Orthogonality exponent for bias quenches in the SBM.--}
\label{sec:exponent-sbm}
Consider the quench in the SBM from a fully biased to an unbiased system. Within the IRLM picture, the initial state Hamiltonian describes a fully occupied ($n_d=1$) resonant level at energy $E=\epsilon=-\infty$ hybridizing with conduction electrons which experience a local potential of strength $+U_{dc}/2$. The resulting conduction electron phase shift at the Fermi level is given by $\delta_i=\arctan(\pi\rho U_{dc}/2)=-\delta$ where $\delta=\arctan(-\pi\rho U_{dc}/2)$ was defined earlier: note also that since the resonant level lies at $E=\epsilon=-\infty$, it makes no contribution to this phase shift. In the final state, the resonant level is switched to $E=\epsilon=0$ following the quench and is therefore half filled ($n_d=1/2$) so the term with $U_{dc}$ is inoperative and the conduction electrons acquire a phase shift due to the resonant level. The final state phase shift of conduction electrons due to this half filled resonant level is then given by the Friedel sum rule $\delta_f=\pi n_d=+\pi/2$. Thus, within the spinless IRLM, we have for the orthogonality exponent 
\begin{align}
\alpha_{\rm OC} &=(\frac{\delta_f-\delta_i}{\pi})^2 = \frac{1}{4}(1 + \frac{2\delta}{\pi})^2=\frac{1}{2}\alpha,\label{eq:alpha_oc_sbm-1}
\end{align}.

For our TDNRG calculations we use $\Lambda=4$, $n_{kept}=860$ kept states per NRG iteration, and Gaussian broadening scheme with broadening constant $\eta=1/32$

\begin{table}[t]
    \begin{tabular}{ccccc}
        \hline
        \hline
        \textbf{\( \Delta_0 \)} &\textbf{\( U_{dc} \)} & \textbf{\( \alpha \)} & \textbf{\( \Delta_r\equiv T_0 \)} & \textbf{\( \beta = 1 - {\alpha}/{2} \)} \\
        \hline
        $0.05$&$-0.377$ & 0.7 & $0.042\times 10^{-3}$ & \(0.65 \) \\
        $0.05$&$0$ & 0.5 & $1.570\times 10^{-3}$ & \(0.75 \) \\
        $0.05$&$0.471$ & 0.3 & $9.677\times 10^{-3}$ & \(0.85 \) \\
       \hline        
        $0.16$&$-0.757$ & 0.9 & $4.672\times 10^{-6}$ & \(0.55 \) \\
        $0.05$&$-0.757$ & 0.9 & $1.828\times 10^{-10}$ & \(0.55 \) \\
        $0.016$&$-0.757$ & 0.9 & $1.938\times 10^{-15}$ & \(0.55 \) \\
        \hline
        $0.05$&$-0.377$ & 0.7 & $0.042\times 10^{-3}$ & \(0.65 \) \\
        $0.005$&$-0.377$ & 0.7 & $1.98\times 10^{-8}$ & \(0.65 \) \\
        $0.0005$&$-0.377$ & 0.7 & $9.089\times 10^{-12}$ & \(0.65 \) \\
        $0.00005$&$-0.377$ & 0.7 & $3.947\times 10^{-16}$ & \(0.65 \) \\
       \hline        
        $0.16$&$-0.658$ & 0.85 & $4.85\times 10^{-5}$ & \(0.575 \) \\
        $0.05$&$-0.658$ & 0.85 & $3.83\times 10^{-8}$ & \(0.575 \) \\
        $0.016$&$-0.658$ & 0.85 & $1.948\times 10^{-11}$ & \(0.575 \) \\
        $0.005$&$-0.658$ & 0.85 & $3.08\times 10^{-15}$ & \(0.575 \) \\
        \hline
        \hline
    \end{tabular}
        \caption{Numerical value of $\alpha$, low energy scale $\Delta_r\equiv T_0$, power-law exponent $\beta=1-\alpha/2$ for different values of $U_{dc}$. $\alpha=\frac{1}{2}(1+\frac{2\delta}{\pi})^2$ and $\delta=\tan^{-1}(-\frac{\pi\rho U_{dc}}{2})$ with the constant density $\rho=1/2D$ and $\Gamma=\pi\rho V^2=0.001D=\pi \Delta_0^2/4\omega_c$ ($\Delta_0\approx 0.05$). The low energy scale of the IRLM is given in terms of the local charge susceptibility via $T_0=1/2\chi(0)$ and is identified with the renormalized tunneling amplitude $\Delta_r$ of the SBM, with $\chi(T)$ calculated from the derivative of occupation w.r.t. bias, $\chi(T) =-\left(\frac{\partial \langle n_d\rangle}{\partial \epsilon_d}\right)_{\epsilon_d=0}$.
    }
    \label{tab:parameters}
\end{table}

\end{acknowledgments}


\bibliography{thermodynamics}
\end{document}